\documentclass[a4paper,pra,twocolumn,notitlepage,superscriptaddress]{revtex4-1}
\usepackage[english]{babel}
\usepackage[utf8]{inputenc}
\usepackage{graphicx,epsfig,color}
\usepackage{latexsym}
\usepackage{amsfonts} 
\usepackage{amsmath}
\usepackage{textcomp}
\usepackage{hyperref}
\usepackage{placeins}

\def\ket|#1>{| #1 \rangle}
\def\bra<#1|{\langle #1 |}
\def\<{\langle}
\def\>{\rangle}
\def\{{\lbrace}
\def\}{\rbrace}
\def\({\left(}
\def\){\right)}
\def\[{\left[}
\def\]{\right]}
\def\beq{\begin{equation}}
\def\eeq{\end{equation}}

\def\R{\mathbb{R}}

\def\nn{\nonumber}

\def\S{{\cal S}}
\def\J{{\cal J}}
\def\I{{\cal I}}
\def\min{\text{min}}

\begin{document}

\title{The hyperlink representation of entanglement and the
  inclusion-exclusion principle}

\author{Silvia N. Santalla} \affiliation{Dto. Física \& GISC,
  Universidad Carlos III de Madrid, Leganés, Spain}

\author{Sudipto Singha Roy}
\affiliation{Department of Physics, Indian Institute of Technology
  (ISM), Dhanbad, India}

\author{Germán Sierra}
\affiliation{Instituto de Física Teórica, UAM-CSIC, Universidad
  Autónoma de Madrid, Cantoblanco, Madrid, Spain}

\author{Javier Rodríguez-Laguna}
\affiliation{Dto. Física Fundamental, Universidad Nacional de
  Educación a Distancia (UNED), Madrid, Spain}

\begin{abstract}
  The entanglement entropy (EE) of any bipartition of a pure state can
  be approximately expressed as a sum of entanglement links (ELs). In
  this work, we introduce their exact extension, i.e. the entanglement
  hyperlinks (EHLs), a type of generalized mutual informations defined
  through the inclusion-exclusion principle, each of which captures
  contributions to the multipartite entanglement that are not
  reducible to lower-order terms. We show that any EHL crossing a
  factorized partition must vanish, and that the EHLs between any set
  of blocks can be expressed as a sum of all the EHLs that join all of
  them. This last result allows us to provide an exact representation
  of the EE of any block of a pure state, from the sum of the EHLs
  which cross its boundary. In order to illustrate their rich
  structure, we discuss some explicit numerical examples using ground
  states of local Hamiltonians. The EHLs thus provide a remarkable
  tool to characterize multipartite entanglement in quantum
  information theory and quantum many-body physics.
\end{abstract}

\date{January 21, 2026}

\maketitle

\section{Introduction}

Fundamental physics has undergone a profound shift in perspective,
elevating information from an abstract concept to a fundamental
physical property in its own right. This paradigm has yielded deep
insights into statistical mechanics, condensed matter physics,
metrology, black hole thermodynamics, and the foundations of
computation \cite{Adesso.19,Mezard.09}. Furthermore, the application
of the concepts from information theory to quantum systems has
provided crucial insights into the nature of entanglement
\cite{Horodecki.09}, as exemplified by the non-positivity of the
quantum conditional entropy \cite{Horodecki.05}.

In \cite{Singha.20} we put forward the {\em entanglement link} (EL)
representation of the entanglement entropy (EE) of all possible
bipartitions of a quantum pure state. Indeed, the EE of any block $A$
can be approximately written as

\beq
S_A \approx \sum_{i\in A,j\in \bar A} J_{ij},
\label{eq:el_rep}
\eeq
where the $J_{ij}\geq 0$ are called ELs. In other terms, the EE of $A$
is approximately computed as the sum of all the ELs which cross the
boundary between $A$ and its complementary $\bar A$, thus presenting a
strong similarity to the {\em area law} of entanglement
\cite{Eisert.10}. Notice that, in an $N$-party system, Eq.
\eqref{eq:el_rep} attempts to reconstruct $O(2^N)$ different EEs using
$O(N^2)$ ELs. Therefore, the accuracy of this approximation is
surprising in practice. In \cite{Singha.21} we provided practical
numerical techniques in order to evaluate the EL of a pure state
expressed in the form of a matrix product state or as a fermionic
Gaussian state, and in \cite{Santalla.23} they were applied to the
evolution of EE after a quantum quench. Moreover, each EL $J_{ij}$ can
be chosen as half the {\em mutual information} between parties $i$ and
$j$ \cite{Singha.20},

\beq
J_{ij}={1\over 2} \I(i,j)
\equiv {1\over 2}\(S_i + S_j - S_{ij}\),
\label{eq:mi}
\eeq
thus showing their non-negativity. We should stress that the EL
representation given by the mutual informations is not optimal, as
discussed thoroughly in \cite{Singha.20}. Interestingly, the common
inequalities regarding EE (subadditivity, strong-subadditivity,
Araki-Lieb, etc.) can be easily proved from the positivity of the ELs.

Yet, we should stress that the EL representation, \eqref{eq:el_rep} is
only approximate. In this work we aim to provide an {\em exact}
extension, associating the EE of any block $A$ to the sum of certain
{\em entanglement hyperlinks} (EHLs) which cross the boundary between
$A$ and its complementary $\bar A$. 

Since the ELs are related to the mutual information, we may conjecture
that the EHLs can be associated to {\em multipartite mutual
  informations}. Yet, the characterization of multipartite
entanglement presents enormous challenges \cite{Ma.23}. Several
definitions have been put forward to this end, each of which is useful
for different applications
\cite{Kumar.17,Kumar.24,Sazim.20,Guo.23,Han.25}. Furthermore,
Meyer-Wallach's global entanglement measure~\cite{Meyer.02}, the
geometric measure of entanglement~\cite{Wei.03}, generalized geometric
measure~\cite{ SenDe.10}, relative entropy of
entanglement~\cite{Vedral.97}, $n$-tangle~\cite{CKW.00}, and squashed
entanglement~\cite{Christandl.04} provide valuable information about
the amount or nature of multipartite entanglement present in a quantum
state. However, these measures often involve optimization procedures
over large sets of quantum states or extensions of the system, which
can become computationally demanding for large many-body systems or
mixed states. In addition to this, the measures often summarize the
entanglement structure into a single scalar quantity and do not reveal
how the correlations are distributed in space or which subsets of
degrees of freedom participate in them. In contrast, the entanglement
hyperlink representation trades compression for interpretability,
providing a detailed picture of the multipartite correlation structure
and identifying the groups of regions that are strongly entangled. We
therefore view EHLs not as a replacement for existing multipartite
entanglement measures, but rather as a complementary tool for
characterizing quantum many-body states.

One of the most popular measures of multipartite information is based
on the {\em inclusion-exclusion principle} from combinatorics
\cite{Allenby.11,Mazur.10}. Unfortunately, it is not consistently
named across the literature, being known as {\em interaction
  information} \cite{McGill.54,Ting.66,Sakaguchi.67} or $I$-measure
\cite{Yeung.07} in classical applications, while in the quantum
information framework it has been called {\em conditional mutual
  information} \cite{Kumar.17} or {\em local quantum information}
\cite{Han.25}, among other denominations. In this work we will call
them EHLs, precisely because they provide our desired extension of the
EL representation.

Classical EHLs have been successfully applied to the statistical
analysis of molecular dynamics simulations \cite{Levine.14},
astrophysical data \cite{Pandey.17} or brain structures
\cite{Varley.23}. Their sign plays a central role in the {\em partial
  information decomposition} from multivariate information theory
\cite{Williams.10,Varley.25} and the associated notions of {\em
  redundancy} and {\em synergy}. In simple terms, three parties show
redundancy when knowing two of them makes the third unnecessary.
Meanwhile, three parties show synergy when the full information about
the system is only achieved when we know all of them, and no two of
them will suffice.

Quantum EHLs of third-order appear in the definition of the
topological EE \cite{Kitaev.06}, and its sign has received a lot of
attention, since the synergetic case translates as {\em monogamy} of
entanglement \cite{Dhar.17}, i.e. the limitation in the ability of a
quantum system to become entangled with many different parties. In the
framework of holographic duality and the Ryu-Takanayagi formula
\cite{Chen.22,Takayanagi.25}, third-order EHLs were proved to be
always monogamous \cite{Casini.09,Hayden.13}, with further studies
conjecturing other limitations in the signs of their higher-order
counterparts \cite{Bao.15,Mirabi.16,Bao.20,Bao.22,Ju.23}. Recently,
the {\em information lattice} framework
\cite{Kvorning.22,Artiaco.25,Bauer.25}, which organizes quantum
information across scales, has been refined via an inclusion–exclusion
formulation \cite{Flor.25}.

This article provides further mathematical insight into the EHLs, with
special emphasis on their use to provide an exact extension of the EL
representation. In Sec. \ref{sec:ehl} we define the EHLs and list some
salient properties, while our main mathematical results and
conjectures are discussed in Sec. \ref{sec:theorems}. Sec.
\ref{sec:numerics} explores numerically the validity of these
properties using some standard quantum pure states. And finally, in
Sec. \ref{sec:conclusions} we summarize our conclusions and proposals
for further work.


\section{Entanglement hyperlinks}
\label{sec:ehl}

The inclusion-exclusion principle is widely employed in many branches
of mathematics, specially in combinatorics \cite{Allenby.11,Mazur.10}.
In its most standard form, we consider $N$ sets, $\{A_i\}$, and assert
that

\begin{align}
  &\left|\bigcup_{i=1}^N A_i\right| =
  \sum_{i=1}^N |A_i| - \sum_{i>j=1}^N |A_i\cap A_j|\; +\nn\\
  +&\sum_{i>j>k=1}^N |A_i \cap A_j \cap A_k| - \cdots + (-1)^{N+1}
  \left|\bigcap_{i=1}^N A_i\right|,
  \label{eq:ie}
\end{align}
i.e. the measure of the union of all the sets is the sum of their
measures, minus the sum of their pair-intersections, plus the sum of
their triple-intersections, etc. For example, for three subsets we
have

\begin{align}
  |A\cup &B\cup C|= |A|+|B|+|C|-\nn\\
  & -|A\cap B|-|A\cap C|-|B\cap C|+|A\cap B\cap C|.
  \label{eq:ie3}
\end{align}
Its main use is to {\em avoid double-counting} in problems related to
combinatorics \cite{Allenby.11}, e.g. the number of {\em derangements}
(permutations which do not leave any element in its original position)
or the evaluation of Euler's $\phi$-function of an integer $n$ (number
of integers with no prime factors in common with $n$).

\medskip

Let us define the entanglement hyperlinks (EHL) of a pure state in
analogy to the {\em inclusion-exclusion} principle,

\begin{equation}
\J_I\equiv \sum_{A\subseteq I} (-1)^{|I|-|A|} S_A,
\label{eq:def_ehl}
\end{equation}
where $I$ is a multi-index denoting a subset of $\Omega$, whose size
$|I|$ will be called the {\em rank}, or number of legs, of the
corresponding EHL. Notice that the sign factor has been changed from
Eq. \eqref{eq:mi}, in order to ensure that the sign of the EE of the
largest block is positive \cite{Note1}. For example, for one, two and
three indices we have

\begin{align}
  &\J_i =  S_i,\nn\\
  &\J_{ij} = S_{ij} - S_i -S_j,\nn\\
  &\J_{ijk} = S_{ijk} - S_{ij} -S_{ik} - S_{jk} + S_i + S_j + S_k,
\end{align}
where $S_i$, $S_{ij}$ and $S_{ijk}$ represent, respectively, EE of
blocks of one, two and three sites. There are $C_{N,k}=N!/(k!(N-k)!)$
rank-$k$ EHLs in a system with $N$ sites, making a total of $2^N$
EHLs, which exceeds the number of different bipartitions for a pure
state. Notice that, by construction, an EHL only depends on the sites
composing its block, and not on the order in which they are written.

\medskip

We may provide a natural definition of {\em conditional} EHLs,

\begin{align}
  \J_{I|J}\equiv &\sum_{A\subseteq I} (-1)^{|I|-|A|}S(A|J) \nn\\
  = &\sum_{A\subseteq
  I} (-1)^{|I|-|A|} \(S(A\cup J) - S(J)\),
\end{align}
which allows us to provide an alternative recursive definition for the
EHL which is common in the literature \cite{Williams.10}. For any site
$i$, we have

\beq
\J_{I\cup i} = \J_{I|i} - \J_I.
\label{eq:recursive_ehl}
\eeq
This recursive definition seems to privilege one of the sites $i$, but
it can be proved that $\J_I$ is independent of the order in which we
{\em grow} the EHL. Moreover, expression \eqref{eq:recursive_ehl}
allows us to suggest an interpretation for the sign of $\J_I$,
somewhat similar to the partial information decomposition
\cite{Williams.10}. Indeed, $\J_I$ measures the amount of correlation
shared by the sites in $I$, and $\J_{I|i}$ measures the amount of
correlation {\em once site $i$ is known}. If knowing $i$ {\em
  increases} the correlation between sites in $I$, we may say that the
system is {\em synergy-dominated}, otherwise we may say that it
presents {\em redundancy-dominated}.

For three parties, synergy implies the negativity of $\J_{ijk}$, which
can be expressed in terms of mutual informations as

\beq
\I(i,j) + \I(i,k) \leq \I(i,jk),
\label{eq:monogamy}
\eeq
which is usually known as a monogamy relation
\cite{Hayden.13,Bao.15,Mirabi.16}, and has been proved in the context
of holographic EE \cite{Casini.09,Hayden.13}.


\section{Theorems}
\label{sec:theorems}

Let us state some useful theorems regarding the EHLs, whose proofs
will be provided in the appendices. Also, let us introduce some
notation. By $I\in A:B$ we will mean that block $I$ crosses the
boundary between $A$ and $B$, i.e.: $I\subseteq A\cup B$, with both
$|I\cap A|$ and $|I\cap B|\neq 0$, while we will assume that $A\cap
B=\emptyset$. This notation may be further generalized as $I\in
A:B:C$, etc.

\subsection{The factorization theorem}
\label{sec:factorization}

EHLs can help detect disentangled states. Indeed, it can be shown that
{\em an EHL vanishes whenever it crosses a partition with zero mutual
  information}. Let us consider $\J_I$, such that $I$ can be
decomposed into two disjoint sets, $I=I_1\cup I_2$. Then,

\beq
\I(I_1,I_2)=0 \quad\Rightarrow\quad \J_I=0,
\eeq
i.e. if the sites forming an EHL can be decomposed into two
complementary blocks with zero mutual information, then the full EHL
is zero. We may prove as a corollary that {\em any EHL crossing a
  partition with zero EE will vanish.} In formal terms, if $I\in
A:\bar A$, i.e. $I$ intersects both $A$ and $\bar A$, then we may
define $I_1=I\cap A$ and $I_2=I\cap \bar A$ and use the previous
theorem in order to show that

\beq
S_A=0 \quad\Rightarrow\quad \J_I=0.
\eeq
Fig. \ref{fig:illust_factorized} provides an illustration. A proof
sketch is provided in Appendix \ref{app:proofs}.

\begin{figure}
  \includegraphics[width=6cm]{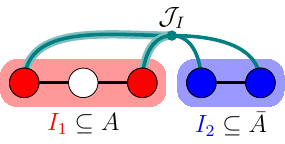}
  \caption{Illustration of the factorization theorem for EHL. If
    $S(A)=S(\bar A)=0$, and $I\in A:\bar A$, then we may define
    $I_1=I\cap A$ and $I_2=I\cap \bar A$ and show that $\I(I_1,I_2)=0$.
    Then, the factorization theorem implies that $\J_I=0$.}
  \label{fig:illust_factorized}
\end{figure}

As a corollary we may prove that all higher-rank EHL for a valence
bond state (VBS) are zero. Indeed, VBS have an exact EL representation
due to their factorizability into pairs of sites. Another interesting
corollary is found if we consider the highest-rank EHL of a quantum
pure state,

\beq
\J_\Omega=\sum_A (-1)^{N-|A|} S_A.
\eeq
Notice that $\J_\Omega=0$ for odd values of $N$, because $S_A$ and
$S_{\bar A}$ appear with opposite signs in the expansion. If $N$ is
even, $\J_\Omega$ is zero if it contains some factorized partition,
i.e. if there is $A\subset\Omega$ such that $S_A=0$. Notice that
converse need not be true: a zero value for $\J_\Omega$ does not imply
the existence of a partition of zero EE.

Continuity arguments suggest that low values of the minimal EE across
all non-trivial partitions in our system, which we may call
$S_\min=\min_{|A|\in \{1\cdots N-1\}} S_A$, will lead to low values of
$\J_\Omega$. Moreover, a similar argument suggests that the
factorization theorem can be extended in order to show that low values
of $\I(I_1,I_2)$ lead to low values of $\J_{I_1\cup I_2}$. These two
claims will be called the {\em factorization conjectures}:

\begin{align}
&\text{Low } S_\min \Rightarrow  \text{ Low }
\J_\Omega, \label{eq:fact1}\\
&\text{Low } \I(I_1,I_2) \Rightarrow \text{ Low } \J_{I_1\cup
  I_2}.\label{eq:fact2}
\end{align}
Numerical exploration provides support to these conjectures, as shown in
Sec. \ref{sec:numerics}.

The previous results suggest that $|\J_A|$ measures the {\em internal
  degree of entanglement} within block $A$. Therefore, monogamy
arguments allow us to conjecture that high values of $|\J_A|$ must be
linked to {\em low} values of $S_A$: 

\beq
\text{Large } |\J_A| \Rightarrow \text{ Low }
  S_A,\label{eq:monoconj}
\eeq
i.e., blocks with large internal entanglement can not achieve large values of entanglement with the rest of the system,
which will be called the {\em monogamy conjecture}, and will also be
checked in Sec. \ref{sec:numerics}.

\subsection{The reconstruction theorems}
\label{sec:reconstruction}

In this section, we discuss how to reconstruct exactly the EE of all
bipartitions from the information contained in the EHL.

\begin{figure*}[ht!]
  \includegraphics[height=6cm]{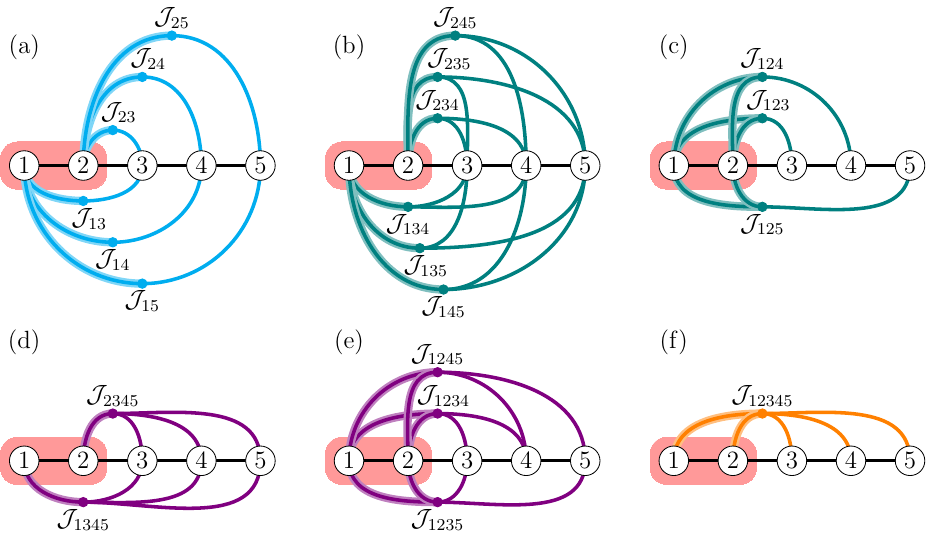}
  \caption{Graphical representation of the edge reconstruction of the
    EE of block $A=\{1,2\}$, Eq. \eqref{eq:edge_recons_i}. The EE of
    block $A$ is (minus one half of) the sum of all the EHLs which
    cross the boundary between $A$ and $\bar A$. We can see (a) the
    two-legged EHLs, (c) and (d) the three-legged EHLs, (d) and (e)
    the four-legged EHLs, (f) the single five-legged EHL (which is
    zero by construction).}
  \label{fig:illust_ehl}
\end{figure*}

\subsubsection{Bulk reconstruction}

The inclusion-exclusion expression can be straightforwardly inverted
using the Moebius formula \cite{Mazur.10}. Let $f:{\cal P}(\Omega)\to
\R$ be a real function over all subsets of $\Omega$, and let us define
another function

\beq
g(A)=\sum_{B\subseteq A} (-1)^{|B|-|A|} f(B),
\eeq
then, we may invert

\beq
f(B)=\sum_{A\subseteq B} g(A).
\label{eq:moebius}
\eeq
Applying this expression to Eq. \eqref{eq:def_ehl} we obtain 

\beq
S_A=\sum_{I\subseteq A} \J_I,
\label{eq:bulk_rec}
\eeq
i.e. the EE of a block $A$ can be obtained adding up all the EHL which
are entirely contained in $A$ \cite{Flor.25}. This expression will be
called {\em bulk reconstruction}, and can be checked simply by
substitution. As a corollary we conclude that {\em the sum of all the
  EHLs of a pure state must be zero}, because they will be equal to
the EE of the full system.

\subsubsection{Edge reconstruction}

Since a pure state has zero entropy, we have for any block $A$ that

\beq
0 = \sum_I \J_I = \sum_{I\subseteq A} \J_I +
\sum_{I\subseteq \bar A} \J_I +  \sum_{I\in A:\bar A} \J_I,
\label{eq:total_decomp}
\eeq
i.e. we have decomposed the set of EHLs into three groups: those
completely contained in $A$, those completely contained in $\bar A$,
and those {\em crossing} the boundary between $A$ and $\bar A$ and
which, therefore, have nonzero intersection with both. Thus, using the
bulk reconstruction expression \eqref{eq:bulk_rec}, we have

\beq
S_A=S_{\bar A}=-{1\over 2} \sum_{I\in A:\bar A} \J_I,
\label{eq:edge_recons}
\eeq
which generalizes the EL representation equation, Eq.
\eqref{eq:el_rep}, with a change in sign. Alternatively, we may write
Eq. \eqref{eq:edge_recons} in a more general way,

\beq
\I(A,\bar A) = -\sum_{I\in A:\bar A} \J_I.
\label{eq:edge_recons_i}
\eeq

\medskip

We should remark that the number of EHLs doubles the number of
different bipartitions. Therefore, they {\em must} contain redundant
information. Appendix \ref{app:evenlegged} shows that, in fact, the EE of
a pure state can be obtained as a linear combination of all the {\em
  even-legged} EHLs, each of them multiplied by a universal prefactor
which only depends on the total number of legs of the EHL and the
number of legs inside the block.

\subsection{The coarse-graining theorem}
\label{sec:coarsegraining}

Eq. \eqref{eq:edge_recons_i} shows that the mutual information between
blocks $A$ and $\bar A$ can be expressed as (minus) the sum of all
EHLs which are contained in $A\cup \bar A$ and cross the boundary
between them, i.e. those presenting non-zero intersection with both.
This structure can be further generalized, as we will show.

Let $A_1\cdots A_K$ be non-overlapping blocks in our system. We may
define a coarse-grained set of EHLs as follows,

\beq
\tilde \J_{A_1\cdots A_K} \equiv \sum_{B\subseteq \{1\cdots K\}} (-1)^{|K|-|B|}
  S\(\bigcup_{i\in B} A_i\),
\eeq
using a notation that we expect to be self-explanatory. We can express
these coarse-grained EHLs in terms of the original EHLs as follows,

\beq
\tilde \J_{A_1\cdots A_K} = \sum_{I\in A_1:\cdots:A_K} \J_I,
\label{eq:coarsegraining}
\eeq
i.e. a coarse-grained EHL is obtained by summing the fine-grained EHLs
with at least one leg on each block and no legs outside the union of
all blocks. Notice that Eq. \eqref{eq:coarsegraining} leads as a
corollary to the edge reconstruction Eq. \eqref{eq:edge_recons_i}.


\section{Numerical examples}
\label{sec:numerics}

In this section, we characterize the EHL of some quantum states of
physical relevance. Concretely, we will focus on some ground states
(GS) of 1D free-fermionic Hamiltonians. The main reason for our choice
is simplicity, since the EE of any block can be easily obtained from
the two-point correlation function \cite{Peschel.03}. Concretely, let
us consider the following free-fermionic Hamiltonian

\beq
H=-\sum_{i=1}^N t_i\, c^\dagger_i c_{i+1} + \text{h.c.},
\eeq
where $c^\dagger_i$ and $c_i$ are canonical fermionic creation and
annihilation operators, such that $\{c_i,c^\dagger_j\}=\delta_{ij}$.
The values $\{t_i\}$ are the hopping parameters, and we will consider
two families:

\begin{itemize}
  \item The dimerized family, $t_i=t_0(1-(-1)^i \delta)$, with
    $\delta\in [-1,1]$. This corresponds to the Su-Schieffer-Heeger
    (SSH) model \cite{Asboth}, which is critical for $\delta=0$ and is
    gapped otherwise.
  \item Random-hopping states, for which $t_i=t_0$ times an i.i.d.
    random variable uniformly distributed in $[0,1]$
    \cite{Ramirez.14b}.
\end{itemize}

In the following, we will consider rather small systems, with up to
$N=10$ sites, because the number of different partitions grows
exponentially with $N$.

\subsection{Factorization properties}

\begin{figure}
  \includegraphics[width=7cm]{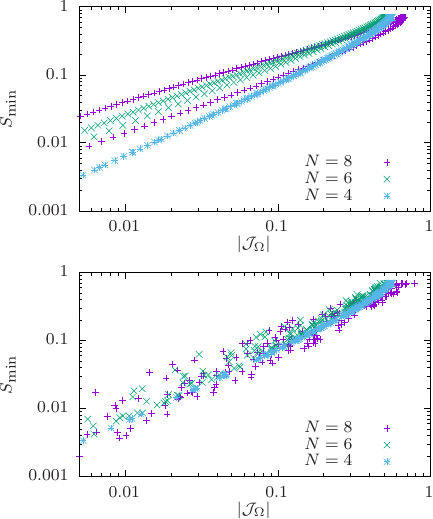}
  \caption{Checking the first factorization conjecture, Eq.
    \eqref{eq:fact1}. Whenever the minimal entropy of a non-trivial
    block $S_\min$ is small, the highest-rank EHL $|\J_\Omega|$ is
    also small, using chains of $N=4$, 6 an 8 sites, (a) Dimerized
    chain, for $\delta\in [-1,1]$; (b) Random chain, using 200
    samples.}
  \label{fig:SminJtot}
\end{figure}

In this section, we will provide numerical evidence in favor of the
factorization conjectures, Eq. \eqref{eq:fact1} and \eqref{eq:fact2},
and the monogamy conjecture, Eq. \eqref{eq:monoconj}. Fig.
\ref{fig:SminJtot} shows the minimal entropy across all the possible
partitions, $S_\min$, against the absolute value of the highest-rank
EHL, $|\J_\Omega|$, for each GS of the considered systems, in
log-scale. Notice the strong nearly linear relation in all cases,
suggesting that the highest-rank EHL correlates with the minimal EE
across all the different bipartitions, as conjectured in Eq.
\eqref{eq:fact1}.

\begin{figure}
  \includegraphics[width=8cm]{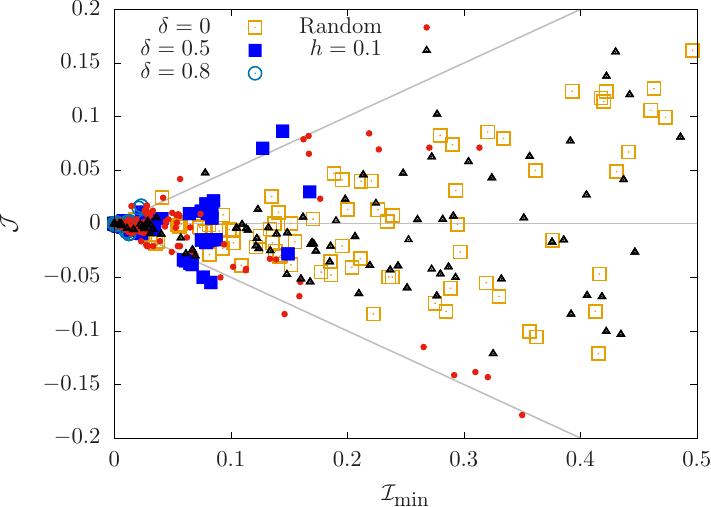}
  \caption{Checking the second factorization conjecture, Eq.
    \eqref{eq:fact2}. We plot $\J_I$ vs the minimal mutual information
    within the block, $\I_\min(I)$ for all blocks of three or more
    legs obtained from some of the states considered in Fig.
    \ref{fig:SminJtot}: dimerized GS with $\delta=0$, 0.5 and 0.8, a
    sample from the random chain always using $N=8$.}
  \label{fig:JJ}
\end{figure}

The second factorization conjecture, Eq. \eqref{eq:fact2}, suggests
that a low value of the minimal mutual information within a block
leads to a low value for the EHL. For all blocks $I$ we have computed
the minimal value taken by the mutual information,

\beq
\I_\min(I) = \min_{I=I_1\cup I_2} \I(I_1,I_2),
\eeq
and plotted these values vs $\J_I$ in Fig. \ref{fig:JJ} for the same
states as in Fig. \ref{fig:SminJtot}. Naturally, we have not
considered EHL with one or two legs, because in that case the relation
becomes trivial.

\begin{figure}
  \includegraphics[width=8cm]{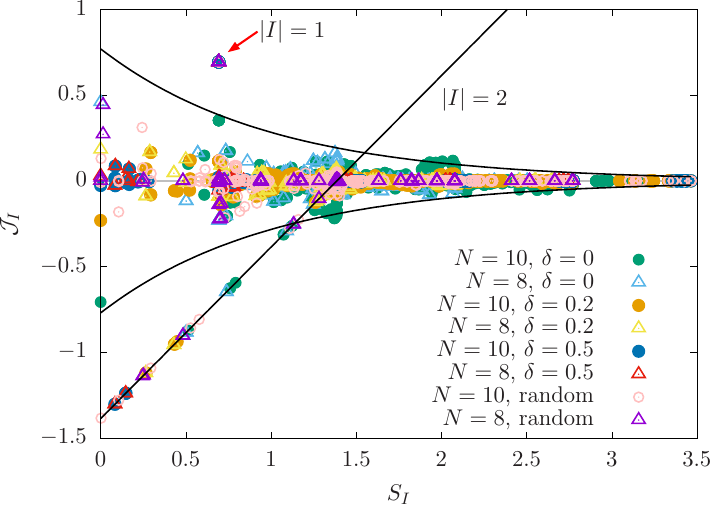}
  \caption{Checking the monogamy conjecture, Eq. \eqref{eq:monoconj},
    showing the EHL vs the EE of all blocks extracted from the
    Gaussian states considered in the previous figures. Notice the
    exponential envelope, which supports our claim that larger EE give
    rise to lower values of the EHL. Blocks with $|I|=1$ and 2 are
    specially marked, since they follow special relations in this
    case.}
  \label{fig:monogamy}
\end{figure}

The previous results point at the idea that the EE and the EHL of a
given block, $S_I$ and $|\J_I|$, play a {\em dual role}: the EE
measures the {\em external} amount of entanglement, while the EHL
measures the {\em internal} amount. The monogamy conjecture, Eq.
\eqref{eq:monoconj} suggests that large values of $S_I$ should lead to
low values of $|\J_I|$. Indeed, Fig. \ref{fig:monogamy} provides some
empirical support for this claim, plotting $S_I$ vs $\J_I$ for all
blocks taken from several of the states previously discussed.
Single-site blocks for these states always present $S_I=\J_I=\log 2$,
and two-site blocks fulfill the exact relation $\J_I=-2\log 2+ S_I$,
and are specially marked in the figure. All higher-rank EHL fall
within the exponential envelope shown in the figure, validating our
conjectured negative monogamous correlation between the EHL and the
EE.

\subsection{EHL signs}

\begin{figure}
  \includegraphics[width=8cm]{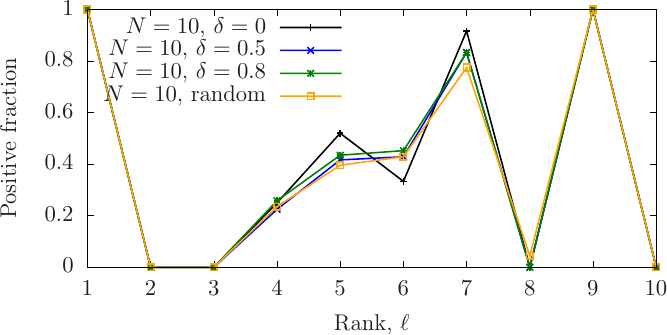}
  \caption{Fraction of EHL which are positive (among those non-zero)
    as a function of their rank, for the aforementioned states with
    $N=10$.}
  \label{fig:signs}
\end{figure}

EHLs of rank 3 should be negative for the entanglement to display
monogamy \cite{Casini.09,Hayden.13}. This was indeed proved within the
holographic framework for EE following the Ryu-Takayanagi formula,
while the extension to higher ranks remains conjectural. Fig.
\ref{fig:signs} shows the proportion of positive EHL (among the
non-zero ones) for the considered states, using $N=10$. For ranks 2
and 3 all the EHL are negative, which is consistent respectively with
the subadditivity of the EE (positivity of the mutual information) and
monogamy. Yet, higher order EHLs can be either positive or negative,
except for very high ranks, which follow a clear alternating pattern.

\subsection{Reconstruction properties}

\begin{figure}
  \includegraphics[width=8cm]{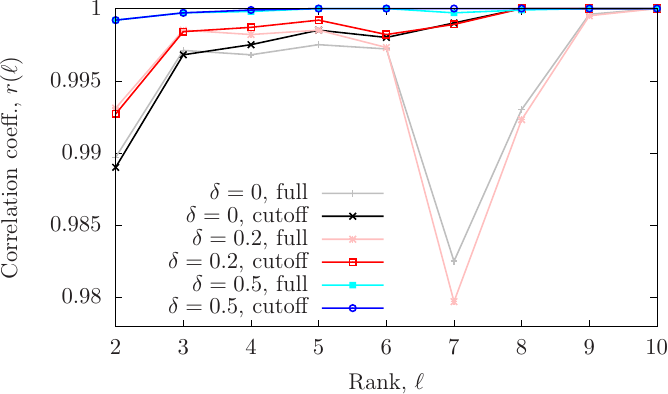}
  \caption{Correlation coefficient $r(\ell)$, defined in Eq.
    \eqref{eq:corr_coef}, between the partial sums $S_A(\ell)$ and
    the exact entropies $S_A$, for different quantum states, as a
    function of the maximal rank $\ell$. Curves labeled as {\em full}
    use the whole set of EE data, while those labeled as {\em cutoff}
    exclude blocks with sizes 1 and 2.}
  \label{fig:errors}  
\end{figure}

The edge reconstruction formula, Eq. \eqref{eq:edge_recons} is an
exact equation, that is fulfilled by all blocks in all possible pure
states. Yet, we may wonder whether partial sums of the terms, obtained
summing up only EHLs up to a certain maximal rank, $\ell$, provide
suitable approximations to the total EE. As we know, the first term
should already be a good approximation, since it is associated to the
EL. Let us define the {\em partial sums}

\beq
S_A(\ell)=-{1\over 2} \sum_{\substack{I\in A:\bar A,\\|I|\leq \ell}} \J_I.
\eeq
In order to quantify the adjustment of the partial sums to the exact
entropies, we use the correlation coefficient, defined as

\beq
r(\ell)\equiv {\sum_A (S_A(\ell)-S(\ell)) (S_A - S) \over
  \sigma(\ell)\,\sigma },
  \label{eq:corr_coef}
\eeq
where $S$ and $\sigma$ are the average and deviation of all $S_A$,
while $S(\ell)$ and $\sigma(\ell)$ are the average and deviation of
all $S_A(\ell)$. Fig. \ref{fig:errors} shows this correlation
coefficient $r(\ell)$, Eq. \eqref{eq:corr_coef}, as a function of the
rank $\ell$, either using all the blocks $A$ or excluding those with
sizes $|A|=1$ and 2, which present a behavior which differs from the
rest, as shown in Fig. \ref{fig:monogamy}. Indeed, the correlation
coefficients when we exclude the small blocks increases almost
monotonously with $\ell$, while for the whole set of EE we find a
sudden decay in the correlation for $\ell=7$. A full explanation for
this this anomalous behavior is not yet known to us, but it is related
to a {\em vertical shift} suffered by the partial sums for certain
block sizes.

\section{Conclusions and further work}
\label{sec:conclusions}

In this work we have explored the physical properties of the
entanglement hyperlinks (EHLs), which are a set of generalized mutual
informations for a multipartite quantum system, defined using the
inclusion-exclusion principle, which have received many different
names in the literature. We have established two factorization
theorems: any EHL will be zero either if it crosses a factorized
boundary, or if it contains a partition with zero mutual information.

Crucially, we have described an exact hyperlink representation of
entanglement, proving that the entanglement entropy (EE) of any block
of a pure state equals (minus a half) the sum of all the EHLs crossing
the boundary between the block and its environment. The hyperlink
representation extends the previously known approximate link
representation, in which the EE of any block was evaluated as the sum
of entanglement links (EL) which cross its boundary. This result is a
corollary of a more general coarse-graining theorem, which states that
the EHL of several blocks can be obtained summing all the EHLs which
cross the common boundary between them.

Finally, we have explored numerically the physical properties of the
EHLs. For a local system, we show that the EHLs provide relevant
information regarding the multipartite entanglement of the block.
Indeed, the EHL becomes zero whenever it crosses a factorized
partition, but continuity ensures that it will still take a small
absolute value when the block is nearly factorized. Also, the value of
an EHL is negatively correlated with its EE, due to entanglement
monogamy.

EHLs are expensive to evaluate, since they require knowledge of the EE
of all possible partitions. Yet, going beyond our current work, they
may provide invaluable insight into the geometrical structure of
entanglement of pure and mixed states. For example, they may point
toward a more precise formulation of a genuinely quantum geometry,
transcending the area-law framework implied by the EL representation.
Additionally, recent efforts to represent quantum states using quantum
circuits have highlighted the importance of identifying the minimal
resources required for state preparation \cite{Zhang.22}. Since
higher-order hyperlinks directly capture genuine many-body
correlations, they may also provide useful insight into the complexity
of constructing a quantum state, for example by estimating the minimum
number of multi-qubit gates needed for its preparation. We plan to
explore this direction in future work.


\begin{acknowledgments}
We acknowledge Olalla Castro-Alvaredo for very interesting discussions.
S.N.S. and J.R.L. acknowledge the Spanish government for financial
support through MINECO grants PID2021-123969NB-I00,
PID2024-159024NB-C21, and PID2024-159024NB-C22 (MCIU/AEI/FEDER, UE).
S.S.R. acknowledges the faculty research scheme at IIT (ISM) Dhanbad,
India under Project No. FRS/2024/PHYSICS/MISC0110, and from the
Anusandhan National Research Foundation (ANRF), Government of In- dia,
under Grants No. ANRF/ARG/2025/004617/PS and No.
ANRF/ECRG/2025/002793/PMS. G.S. acknowledges financial support from
the Spanish MINECO grant PID2021-127726NB-I00, and
PID2024-161474NB-I00 (MCIU/AEI/FEDER, UE), and from QUITEMAD-CM
TEC2024/COM-84. We also acknowledge support from the Grant IFT Centro
de Excelencia Severo Ochoa CEX2020-001007-S funded by
MCIN/AEI/10.13039/501100011033, and from the CSIC Research Platform on
Quantum Technologies PTI-001 and the QUANTUM ENIA project Quantum
Spain funded through the RTRP-Next Generation program under the
framework of the Digital Spain 2026 Agenda.
\end{acknowledgments}


\appendix

\onecolumngrid

\section{Proof sketches for the theorems}
\label{app:proofs}

In this appendix we provide proof sketches for the theorems stated in
the main text, along with further examples.

\subsection{Factorization theorems}

As it was discussed in Sec. \ref{sec:factorization}, an EHL vanishes
whenever it crosses a factorized boundary. First of all, we should be
aware that, if $S_A=0$, then for any other block $B$ we have

\beq
S_B=S_{B\cap A} + S_{B \cap \bar A}.
\eeq
Let us now consider an EHL $\J_I$ which crosses the boundary between
$A$ and $\bar A$, i.e. $I$ has non-empty intersections both with $A$
and $\bar A$. We need to prove that $\J_I=0$. Indeed, we have

\begin{align}
\J_I=(-1)^{|I|} \sum_{B\subseteq I} (-1)^{|B|} S_B = (-1)^{|I|}
\sum_{B\subseteq I} (-1)^{|B|} \(S_{B\cap A} + S_{B\cap \bar A}\).
\end{align}
We can disregard the global sign $(-1)^{|I|}$, and notice that $\J_I$
is built of two sums, one composed of EE {\em living entirely in $A$}
and the other composed of EE {\em living entirely in ${\bar A}$}. Let
us focus on the first:

\beq
\sum_{B \subseteq I} (-1)^{|B|} S_{B\cap A}.
\label{eq:partial_sum}
\eeq
Each EE in this sum appears several times, but with different signs.
Let $B=C_1 \cup C_2$, with $C_1\subseteq A$ and $C_2\subseteq \bar A$.
The EE $S_{C_1}$ appears in \eqref{eq:partial_sum} once for each
possible $C_2$, which is the power set of $\bar A\cap I$, $2^{|\bar
  A\cap I|}= 2^{2l-p}$. Out of this set, half of the terms will have a
positive sign and half a negative sign, due to basic properties of the
combinatorial numbers, so the total contribution is zero.

\medskip

Let us provide an example of this result for $N=4$ parts, such that
$S_{12}=S_{34}=0$, and evaluate $\J_{1234}$:

\begin{align}
  \J_{1234}=&S_{1234}-S_{123}-S_{124}-S_{134}-S_{234}+S_{12}+S_{13}+S_{14}+S_{23}+S_{24}
  +S_{34}-S_1-S_2-S_3-S_4\nn\\
  =&(S_{12}+S_{34})-(S_{12}+S_3)-(S_{12}+S_4)-(S_1+S_{34})-(S_2+S_{34})+\nn\\
  &+S_{12}+(S_1+S_3)+(S_1+S_4)+(S_2+S_3)+(S_2+S_4)+S_{34}-S_1-S_2-S_3-S_4=0.
\end{align}

A word of caution is required: the presence of a factorized boundary
in a block ensures that the corresponding EHL will be zero, but the
converse statement is not true. We may have a zero EHL without a
factorized boundary.

\subsection{Coarse-graining theorem}

As claimed in the main text, the EHL of a set of coarse-grained blocks
within the system may be obtained summing all EHLs of the fine-grained
partition which have at least one leg in each of the coarse-grained
parties, Eq. \eqref{eq:coarsegraining}. Again, this is merely a
consequence of the bulk-reconstruction formula, i.e., Moebius
inversion formula \cite{Mazur.10}. Let us prove our theorem for two
and three blocks, hoping that the general trend will be manifest by
then.

Let us consider two blocks, $A$ and $B$. The coarse-grained EHL
connecting them is

\beq
\tilde\J_{AB}=S(A \cup B)-S(A)-S(B)=\sum_{I\subseteq A \cup B} \J_I
- \sum_{I\subseteq A} \J_I - \sum_{I\subseteq B} \J_I,
\eeq
where we have made use of the bulk reconstruction expression. Notice
that the set of EHLs in $A\cup B$ can be classified into three groups:
those which belong to $A$ only, those which belong to $B$ only, and
those which cross the $A:B$ boundary. Thus, we have

\beq
S(AB)=\sum_{I\subseteq A} \J_I + \sum_{I\subseteq B} \J_I + \sum_{I\in A:B}
\J_I,
\eeq
using our notation $I\in A:B$ for those EHLs which are contained in
$A\cup B$ and overlap both $A$ and $B$. From this equation, the
theorem becomes manifest,

\beq
\tilde\J_{AB} = \sum_{I\in A:B} \J_I.
\eeq

Now, let us consider the case with three blocks, $A$, $B$ and $C$,

\beq
\tilde\J_{ABC}=S(A\cup B\cup C)-S(A\cup B)-S(A\cup C)-S(B\cup C)
+S(A)+S(B)+S(C).
\eeq
In this case, we may decompose the set of all EHLs in $A\cup B\cup C$
as belonging to one and only one of the following seven groups,

\begin{itemize}
\item EHLs belonging entirely to $A$, or $B$ or $C$.
\item EHLs crossing the boundary $A:B$ (and not touching $C$), and
  their counterparts regarding $A:C$ and $B:C$.
\item EHLs which cross the triple boundary, $A:B:C$.
\end{itemize}

Thus, we write

\beq
S(ABC)=\(\sum_{I\in A:B:C} + \sum_{I\in A:B} + \sum_{I\in A:C} +
\sum_{I\in B:C} + \sum_{I\subseteq A} + \sum_{I\subseteq B} + \sum_{I\subseteq C}\)
\J_I.
\eeq
Now, just a simple application of the inclusion-exclusion principle
shows the desired result,

\beq
\tilde\J_{ABC}=\sum_{I\in A:B:C} \J_I.
\eeq
This result carries over to any number of parts, and the proof is not
hard to write, even if it may become cumbersome without a proper
notation.


\section{Even-legged edge reconstruction}
\label{app:evenlegged}

\begin{figure*}[ht!]
  \includegraphics[height=3.5cm]{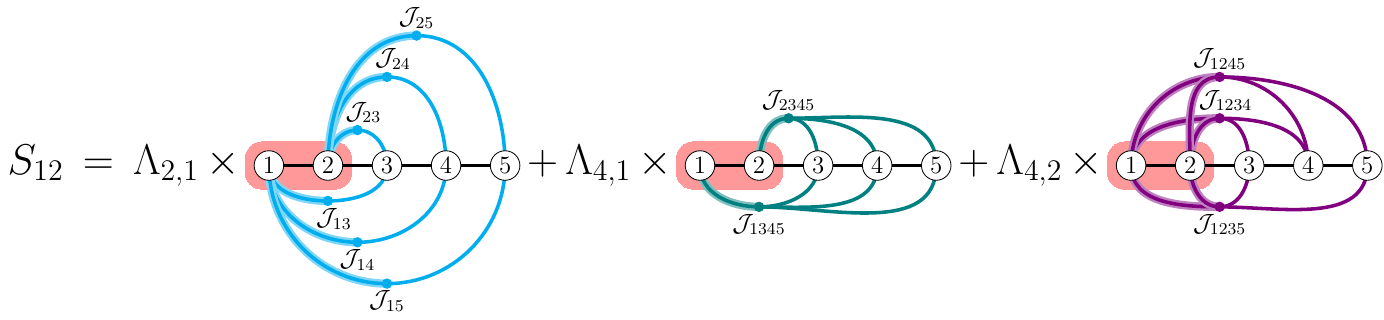}
  \caption{Graphical representation of the edge reconstruction of the
    EE of block $A=\{1,2\}$, Eq. \eqref{eq:edge_even_rec}. The first term
    comprises all the two-legged EHL (aka the EL), with their
    prefactor $\Lambda_{2,1}$, because the have a single leg inside
    the desired block. The second term contains the four-legged EHL
    which have a single leg inside the block, with their prefactor
    $\Lambda_{4,1}$. The third and last term contains the four-legged
    EHL which have two legs inside the block, with their prefactor
    $\Lambda_{4,2}$. The numerical values of the prefactors are given
    in Eq. \eqref{eq:leg_factors}.}
  \label{fig:illust_even_ehl}
\end{figure*}

Let us notice that the number of EE and the number of EHL is exactly
the same, $2^N$. Yet, for a pure state, the number of {\em
  independent} EE is reduced to $2^{N-1}-1$. Therefore, the set of
EHLs can not be independent. We are thus led to consider the
possibility that only a subset of the EHL is needed for the
reconstruction. Since we plan to extend the EL representation, we use
the identity

\beq
\sum_{m=1}^{N/2} C_{N,2m} = 2^{N-1}-1,
\eeq
in order to show that the number of {\em even-legged} EHL coincides
with the number of independent EE in a pure state. Thus, we may
conjecture that even-legged EHL suffice in order to reconstruct the
EE. Let us consider the first examples:

\begin{itemize}
  \item For $N=3$, there are 3 EE, and 3 rank-2 EHL.
  \item For $N=4$, there are 7 EE, 6 rank-2 and 1 rank-4
    EHL, so 7=6+1.
  \item For $N=5$, there are 15 EE, 10 2-leg and 5 4-leg EH,
    so 15=10+5.
  \item For $N=6$, there are 31 EE, 15 rank-2, 15 rank-4 and 1 rank-6
    EHL, so 31=15+15+1.
\end{itemize}

Let us put forward our proposal for the reconstruction formula,

\beq
S_A=\sum_{|I| \text{ even}} \Lambda_{|I|,|I\cap A|}\, \J_I,
\label{eq:edge_even_rec}
\eeq
i.e. we sum over all EHL, each of them multiplied by a {\em
  leg-factor} $\Lambda_{2l,p}$, which depends on the rank of the EHL,
$|I|=2l$ and the number of legs that the EHL has in block $A$,
$p=|I\cap A|$. If $p=0$ or $p=2l$ then $\Lambda_{2l,p}=0$, implying
that we only need to consider EHL which {\em cross the boundary}
between $A$ and $\bar A$. Alternatively, we can say

\begin{equation}
S_A=\sum_{l=1}^{\lfloor {N\over 2} \rfloor} \sum_{p=1}^{2l-1}
\Lambda_{2l,p} \sum_{|I|=2l, |I\cap A|=p} \J_I.
\label{eq:edge_rec_2}
\end{equation}
Of course, expessions \eqref{eq:edge_even_rec} and
\eqref{eq:edge_rec_2} remain a conjecture at this stage. It is still
to be proved that the leg-factors $\Lambda_{2l,p}$ exist and do not
depend on $N$. If they exist, then complementarity symmetry imposes
that $\Lambda_{2l,p}=\Lambda_{2l,2l-p}$. In the next paragraphs we
will determine the leg-factors $\Lambda_{2l,p}$.

\medskip

Let us provide a complete example, based on the illustration shown in
Fig. \ref{fig:illust_even_ehl}. Consider a system with $N=5$ parties, then
the EE for any block can be built in the following way,

\begin{align}
  & S_1 = \Lambda_{2,1} \(\J_{12}+\J_{13}+\J_{14}+\J_{15}\)
  + \Lambda_{4,1} \(\J_{1234} + \J_{1235} +\J_{1245} +\J_{1345}\),\nn\\
  & S_{12} = \Lambda_{2,1}
  \(\J_{13}+\J_{14}+\J_{15}+\J_{23}+\J_{24}+\J_{25}\)  + \Lambda_{4,1}
  \( \J_{1345} + \J_{2345} \) + \Lambda_{4,2} \(\J_{1234} + \J_{1235}
  + \J_{1245}\),\nn\\
  & S_{123} = \Lambda_{2,1}
  \(\J_{14}+\J_{15}+\J_{24}+\J_{25}+\J_{34}+\J_{35}\)+ \Lambda_{4,2}
  \(\J_{1245}+\J_{1345}+\J_{2345}\) + \Lambda_{4,3}
  \(\J_{1234}+\J_{1235}\).\nn\\
  & S_{1234} = \Lambda_{2,1}
  \(\J_{15}+\J_{25}+\J_{35}+\J_{45}\) +
  \Lambda_{4,3}\(\J_{1235}+\J_{1245}+\J_{1345}+\J_{2345}\). 
\end{align}
In order to obtain the leg-factors, we {\em invert} this expression
and read the coefficients. This way we obtain, for the smallest sizes,

\begin{align}
  \Lambda_{2,1}&=-\frac{1}{2}, \nn\\
  \Lambda_{4,1}&=\frac{1}{4},  &\Lambda_{4,2}&=\frac{1}{2}, \nn\\
  \Lambda_{6,1}&=-\frac{1}{2}, &\Lambda_{6,2}&=-1,
  &\Lambda_{6,3}&=-\frac{5}{4}, \nn\\
  \Lambda_{8,1}&=\frac{17}{8},  &\Lambda_{8,2}&=\frac{17}{4},
  &\Lambda_{8,3}&= \frac{47}{8}, &\Lambda_{8,4}&=\frac{13}{2}.
  \label{eq:leg_factors}
\end{align}
Notice that $\Lambda_{2l,p}$ is negative whenever $l$ is odd. In
practice, the leg-factors are obtained by induction, requiring
consistency between the decomposition and the reconstruction formulas
for successive values of $N$. New values of $\Lambda_{2l,p}$ appear at
even values of $N$, which are can be obtained solving a simple linear
equation. At odd values, $N+1$, the same values are confirmed, and
they can be considered fixed for larger values of $N$.

Notice that the reconstruction formula, \eqref{eq:edge_even_rec},
implies that if all the EHL across a boundary are zero, then the EE
associated must also be zero. We should stress that the validity of
Eq. \eqref{eq:edge_even_rec}, with leg-factors that do not depend on
$N$, has only been checked up to $N=9$, and remains conjectural for
larger sizes.

\medskip

Both the set of EE and the set of EHL, $\S$ and $\J$, can be regarded
as a mapping from the power set of $\Omega$ into the real numbers. The
definition of the EHL, Eq. \eqref{eq:def_ehl}, provides a linear
mapping from $\S$ to $\J$, which can be expressed in matricial form,
$\J=M\S$, with $M$ a linear operator in $\R^{2^N}$. Conversely, for a
pure state, $\S$ can be reconstructed either the bulk reconstruction,
the edge reconstructions, or the even-legged edge reconstruction, Eqs.
\eqref{eq:bulk_rec}, \eqref{eq:edge_recons} and
\eqref{eq:edge_even_rec}, respectively. Since all three reconstruction
operations are again linear, we may claim that Eq. \eqref{eq:def_ehl}
has {\em several different inverses}. This apparent contradiction is
due to the fact that we are considering vectors $\S$ with a special
symmetry, i.e.: those corresponding to pure states, in which
$S_A=S_{\bar A}$.

Notice that the existence of the even-legged edge reconstruction
formula implies that, for a pure state, odd-ranked EHLs can be
obtained from the even-ranked ones. I.e. since the set of EE is not
independent, neither is the set of EHL.

\twocolumngrid

\end{document}